\documentclass[aps,prl,preprint,floats,showpacs]{revtex4}
\usepackage{graphicx}
\usepackage{subfigure}
\usepackage{epsfig}
\usepackage{color}
\usepackage{ulem}

\newcommand{\w}{{\omega}}

\renewcommand{\l}{{\lambda}}
\newcommand{\G}{{\Gamma}}

\newcommand{\W}{{\Omega}}

\newcommand{\e}{{|e\rangle}}
\newcommand{\g}{{|g\rangle}}
\newcommand{\s}{{|s\rangle}}

\begin{document}

\title{Electromagnetically Induced Transparency Spectroscopy.}
\begin{abstract}

We propose a method based on the
Electromagnetically Induced Transparency (EIT) phenomenon
for the detection of
molecules which exist as a small minority in the presence of
a majority of absorbers.
The EIT effect we employ effectively eliminates
the absorption of the majority species
in the spectral region where it overlaps with the absorption of the minority species.
The method can also be used
to enhance local-modes transitions which overlap spectrally
with a background of other local-modes transitions of the same molecule.
The general theory is applied to the case of sparse
and congested background spectra within the same molecule
and to the recording of
the spectra of isotopomers (of Chlorine and Methanol) that are in minority
relative to other isotopomers which constitute the majority of molecules
present.
\end{abstract}

\date{\today}
\author{Asaf Eliam$^1$, Evgeny A.~Shapiro$^1$, Moshe Shapiro$^{1,2}$,
~\\\it Departments of Chemistry$^1$ and Physics$^2$,
The University of British Columbia
\\  2036 Main Mall, Vancouver, BC, Canada V6T 1Z1}

\maketitle

\section*{1. INTRODUCTION}

\begin{figure}[ht]
\hskip 0.2 truein
\includegraphics[width=0.8\columnwidth]{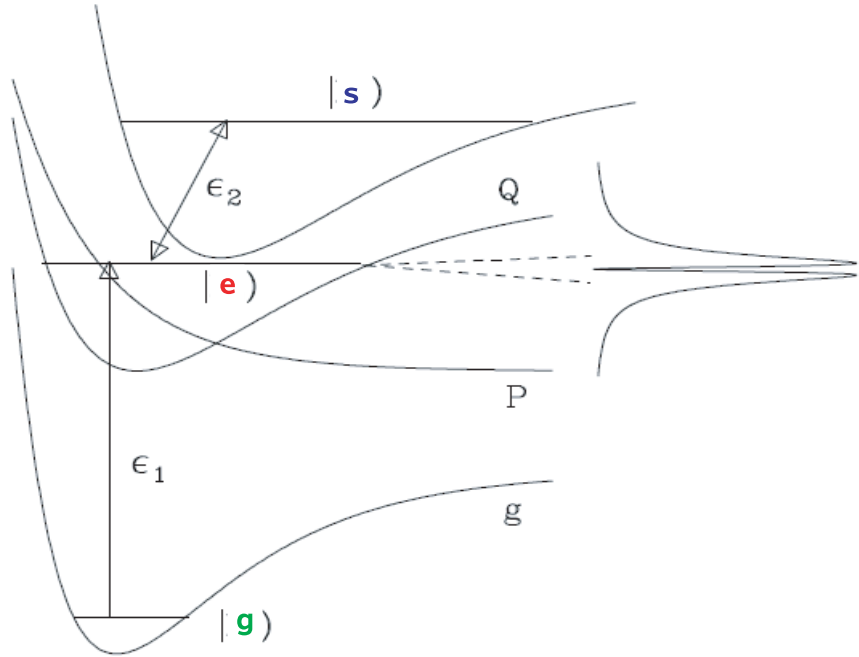} \centering
\vskip 0.2 truein \caption{Molecular EIT: an initial state
$|g\rangle,$ residing in the ground electronic potential is
excited by a weak laser pulse $\epsilon_1$ to a resonance state
$|e\rangle$ in electronic state $Q,$ which decays radiatively or
non-radiatively to the continuum manifold of electronic state $P$.
The $|e\rangle$ state is coupled optically to a third state
$|s\rangle$ by a strong control field $\epsilon_2$ and undergoes
as a result Autler-Townes splitting. As a result of the splitting
and the decay, an EIT ``hole" in the absorption of the probe field
$\epsilon_1$ is formed at $E=E_e$.} \label{reson}
\end{figure}

Electromagnetically Induced Transparency
(EIT)\cite{Harris1997,Boller1991,Field1991,fleishhauer05}
is usually observed in
three-level atomic systems interacting
with two (``probe'' and ``control'') laser fields.
The probe's electric field,
denoted, as illustrated in Fig. \ref{reson},
by $\epsilon_1,$ is in resonance with the transition frequeyncy between
the ground state $|g\rangle$ and an excited state $|e\rangle$.
The ``control'' laser, whose electric field is denoted as
$\epsilon_2,$ couples the excited state $|e\rangle$ to an
auxiliary state $|s\rangle.$ When the conditions are right,
quantum interference stabilizes the ground state
population, causing the transition
from the ground state $|g\rangle$ to the excited state $|e\rangle$
to disappear, rendering the medium {\it transparent}
to the probe field.

An important element at the heart of the EIT effect is that the
excited state $|e\rangle$ is coupled to a (radiative or
non-radiative) continuum, forming a (scattering) resonance. As
depicted schematically in Fig. \ref{reson}, when such a scattering
resonance is dressed radiatively by the control field, it splits
into two (Autler-Townes) components\cite{Autler1955}, forming two
overlapping scattering
resonances\cite{Harris1997,Fano1961,Shapiro1972,G.Kurizki1989}
which interfere destructively at the position of the bare state
energy\cite{Shapiro1972,Shapiro2007}.

This interference results in the appearance of a ``dark
resonance''\cite{Shapiro1972} or a ``dark
state''\cite{Arimondo1976,Arimondo1996} (DS), leading to
``Coherent Population Trapping''
(CPT)\cite{Arimondo1976,Arimondo1996} and transparency.

In the general framework of
nonlinear optics\cite{N.Bloembergen1987,Clark1988,Brumer86},
EIT is used in the
slowing down\cite{Harris1997,Lukin1997,Hau1999} and
storage\cite{Hau2001,Walsworth2001,Matsko2001,Dey2003,Fleischhauer2002}
of
light, in quantum
teleportation and communication\cite{Duan2001}, in high precision
magnetometry\cite{Schwindt2004}, and in developing frequency
standards\cite{Affolderbach2005}. Related interference phenomena are
Electromagnetically Induced
Absorption(EIA)\cite{Lezama1999,Taichenachev1999} and
``Laser Induced Continuum Structures'' (LICS)
\cite{lics,knight,charalambidis,halfmann}.



{The EIT effect is almost insensitive to spontaneous emission from
the excited intermediate state, since the population is trapped in
the two ground states. Moreover,
the transparency window in the absorption spectrum can be
controlled via the control of the
Autler-Townes(AT)\cite{Autler1955}
splitting on top of the
destructive interference of the two overlapping resonances.
These characteristics make EIT a promising tool for high
resolution spectroscopy. Indeed, EIT has been used in the
spectroscopic studies \cite{Lukin1997,Taichenachev1999,Goren2004}
of alkali atoms, mainly for
reducing the inhomogeneous line-widths of absorption lines.

In the present paper we develop EIT as a tool for distinguishing
between overlapping transition lines in atomic and molecular systems.
%
An overlap of two or more transition lines in a mixture of different molecules
(the ``intermolecular'' overlap), or in two (local or collective) modes of the
same molecule (the ``intramolecular'' overlap) is a common
occurrence\cite{Hessel1974,Caldwell1980,Coxon1978,G.1994}. Such an overlap
can reduce our ability to detect molecules in small concentrations in the
presence of molecules in higher concentrations,
reduce the accuracy of line assignments, and lead to errors in the
derived molecular structure. Line overlaps can also generate a
''chain''of misassignments\cite{G.1994},
in which errors in the assignments of some lines can lead to the
misinterpretation
of a whole series, even when it composed of well separated lines.

Commonly used
solutions of this problem utilize numerical fitting algorithms.
However, numerically distinguishing between spectral lines is not
possible when the lines overlap too closely. Likewise, numerical fittings become
ineffective
when one probes a mixture of different molecules with
spectral transitions at similar frequencies. An illustration
is given in Fig.~\ref{fig12}.

\begin{figure}[ht!]
\includegraphics[width=0.7\columnwidth]{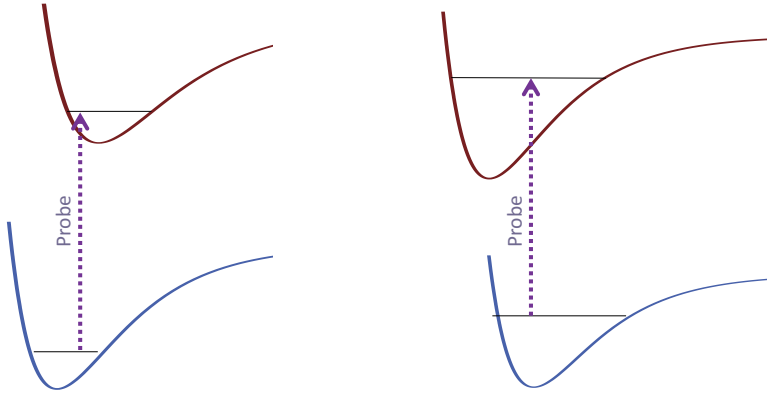} \centering
\caption{(Color online) Two overlapping absorption lines related
to two different vibronic transitions in two molecules.  }
\label{fig12}
\end{figure}

The present proposal uses EIT to overcome the overlap problem
in both the intramolecular and intermolecular regimes.
This is done by applying a ``control'' field which
couples the strongly absorbing states which mask the weakly absorbing lines to
some auxiliary states that do not exist (or are out of resonance)
in the weakly absorbing species.
By making the strongly absorbing lines transparent, the EIT effect helps
accentuate the weakly absorbing features, which now
stand out against a zero
background.
In this way, only the desired spectral lines that have
not been made transparent remain in the spectrum. The details of the
scheme, outlined in Fig.2, are discussed in the next Section.

The structure of this paper is as follows: In section 2 we describe
the theory of EIT spectroscopy. In section 3 we discuss the range of
applicability of the method and in section 4 we demonstrate the viability
of the method by deciphering the spectra in a number of cases
where spectral overlaps occur, such as
a mixture of Cl$_2$ isotopomers and in the spectrum of
the CH$_3$OH molecule. In section 5 we
discuss the merits of the method and compare it to
methods derived from Coherent Anti-Stokes
Raman Spectroscopy (CARS) for achieving the same goals.

\section*{2. THEORY AND THREE SCENARIOS}

\noindent
{\bf 2.1. EIT line shapes}

\noindent
{{We begin by outlining the theory of EIT for strongly decaying
intermediate levels
\cite{Shapiro2007,Shapiro2002}. The consideration proceeds in
three steps. First, one accounts for the states $\s$ and $\e$ in
the three-level configuration depicted in Fig. \ref{reson}. These
states are coupled by the control field characterized by center
frequency $\w_c$, amplitude $\epsilon_c$, and Rabi frequency $\W_c
= \mu_{e,s} \epsilon_c$. Here $\mu_{e,s}$ are the transition-dipole
matrix elements between the $\e$ and $\s$ molecular states.

Expressed in the appropriate co-rotating frame, the
Hamiltonian within the Rotating Wave approximation is given in
atomic units ($\hbar=1$), as,
\begin{equation}
H_{e,s}=  %
\delta/2\left|s\right\rangle\left\langle s\right|- \delta/2
\left|e\right\rangle\left\langle e\right|+\Omega_{c}
\left|e\right\rangle\left\langle s\right|+\Omega_{c}^*
\left|s\right\rangle\left\langle e\right|  \label{RWA}
\end{equation}
where  $\delta=\w_c-(E_e-E_s)$ is the detuning of the control
field from resonance, $E_e$ and $E_s$ are the bare-states energies.
The eigenvalues of this system, given as,
\begin{equation}
\l_j = (3-2j)\W' = (3-2j)\sqrt{\delta^2/4+|\Omega_c|^2},~~~j=1,2,
\label{ATenergies}
\end{equation}
reveal AT split\cite{Autler1955} spectra. In the original
Schroedinger frame, the corresponding eigenstates are given
as\cite{Shapiro2007},
\begin{equation}
\left|\lambda_{j}(t)\right\rangle=\Big(\alpha_{j}\left|s\right\rangle
 e^{i(E_e-E_s+\delta)t}
+\beta_{j}\left|e\right\rangle \Big)
\,e^{-i(\lambda_{j}+\delta/2+E_e)t},~j=1,2,
\end{equation}
where
%
\begin{eqnarray}
\alpha_{j}&=&\left(\frac{\left|\Omega\right|^2}{2\Omega'^{2}-(3-2j)\delta\Omega' }\right)^{1/2},~~~{\rm  and~}~~~\nonumber\\
\beta_{j}&=&\left(\frac{\left|\Omega\right|^2}{2\Omega'^{2}-(3-2j)\delta\Omega' }\right)^{1/2}\frac{\Omega'-(3-2j)\delta/2}{\Omega^{*}}
,~j=1,2. \label{beta} 
\end{eqnarray}

\vskip .1truein

In the second stage of the development we switch on $H_{b-c},$ the bound-continuum
interaction between state $|e\rangle$ and a
continuum of incoming scattering states,
$|E,n_{1}^{-}\rangle,$
where $E$ is  the energy and $n_1$ -
the ``channel index'' - stands for the collection of
all the discrete quantum numbers necessary to fully characterize the
continuum. The $n_1$ subscript is a reminder that the
$|E,n_{1}^{-}\rangle$ states diagonalize only part of the Hamiltonian,
the part which does not contain state $|e\rangle$.

In the absence of a control field
and when the coupling to the continuum is relatively independent of $E$,
state $\e$ decays at a rate given by $\G =  \pi |V_{e,E}|^2$ \cite{Shapiro2002}
where $V_{e,E}\equiv\langle e | H_{b-c} | E, n_1^- \rangle.$
The bound-continuum coupling turns each dressed state
$|\lambda_{1,2}\rangle$ into a fully interacting
scattering resonance $|E,n^-\rangle,$ possessing
both a bound ($|\lambda_{1,2}\rangle$) component
and a continuum ($|E,n_1^-\rangle$)
component. Explicit expressions for these eigenstates, found by
applying overlapping resonances Feshbach's or Fano's
partitioning techniques\cite{Fano1961,Shapiro1972},
can be found in Refs.\cite{Shapiro2002,Shapiro2007}.

At the third stage, we add a ``probe'' laser, whose
electric field is denoted as $\epsilon_{pr},$ to this system.
The latter couples the ground state $\g$ to
the continuum of eigenstates $|E,n^-\rangle$. The one-photon
absorption probability for a weak probe is given in first
order perturbation theory as
\begin{equation}
P_{pr}(E)=2\pi
\left|\epsilon_{pr}(\omega_{E})\mu_{E,g}\right|^{2},
\label{P_pr}
\end{equation}
where $E,$ the scattering energy, satisfies the resonance condition, %
\begin{equation}E = E_g + \hbar\w_E ~, %
\end{equation}
with $\w_E$ being the probe center-frequency.  $\mu_{E,g}$
are the transition-dipole matrix elements,
\begin{equation}
\mu_{E,g} = \langle E,n^{-} |\mu |g\rangle = \sum_{j=1,2} \langle
E,n^{-}|\lambda_{j}\rangle \langle\lambda_{j}| \mu |g\rangle.
\label{mu}
\end{equation}
%
%
The $\langle E,n^{-}|\lambda_{i}\rangle$ overlap integral can be
calculated\cite{Shapiro2007} as,
\begin{equation}\label{bound_contin}
\langle E,n^{-} |\lambda_{i} \rangle  =
\sum_{j}  \langle E, n_1^- | H_{b-c} | \lambda_{j} \rangle
\langle \lambda_{j} |\mathbf{H}(E)^{-1} |\lambda_{i} \rangle,
\end{equation}
where $\mathbf{H}(E)$ is an {\it effective coupling Hamiltonian} describing
the continuum-mediated interaction between the
field-dressed eigenstates. In the case of an
``unstructured'' continuum, where
$\langle\lambda_{j} | H_{b-c}| E,n_1^-\rangle =
\sqrt{\Gamma/\pi}$, $\mathbf{H}(E)$ can be expressed as
\begin{equation}
\mathbf{H}(E)=\left(
\begin{array}{cc}
E-E_{e}-\delta/2-\lambda_{1}-i\Gamma/2    & -i\Gamma/2  \\
-i\Gamma/2                                 & E-E_{e}-\delta/2-\lambda_{2}-i\Gamma/2 \\
\end{array}
\right). \label{hamil}
\end{equation}
The diagonal matrix elements of Eq. (\ref{hamil})
embody the AT level splitting, and the off-diagonal elements describe
the coupling to the continuum.

Equations (\ref{bound_contin},\ref{hamil}) lead to the final
expression
\begin{equation}
\mu_{E,g}=V_{E,e}\left\{\left|\alpha_{1}\right|^{2}D_{11}+
\alpha_{1}\alpha_{2}^{*}D_{12}+\alpha_{2}\alpha_{1}^{*}D_{21}+
\left|\alpha_{2}\right|^{2}D_{22}\right\},
\label{mu_nonres}
\end{equation}
where $D_{ij}=\mathbf{H}^{-1}_{i,j}$, and $\alpha_{j}$ is given by
Eq.(\ref{beta}). If the control detuning $\delta=0$, then
\begin{equation}\label{dipole_AT}
\mu_{E,g}=\frac{1}{D}\mu_{e,s} V_{E,e}[E-E_{e}], \label{mu_res}
\end{equation}
where
\begin{equation}
D=[E-E_{g}-i\Gamma/4]^{2}+\G^2/16-\W_c^{2}.
\end{equation}
Expressions (\ref{P_pr},\ref{mu_nonres},\ref{mu_res}) will be used
below for calculating the field-modified absorption spectra of
molecules.

\begin{figure}[ht!]
\centering
\includegraphics[width=0.8\columnwidth]{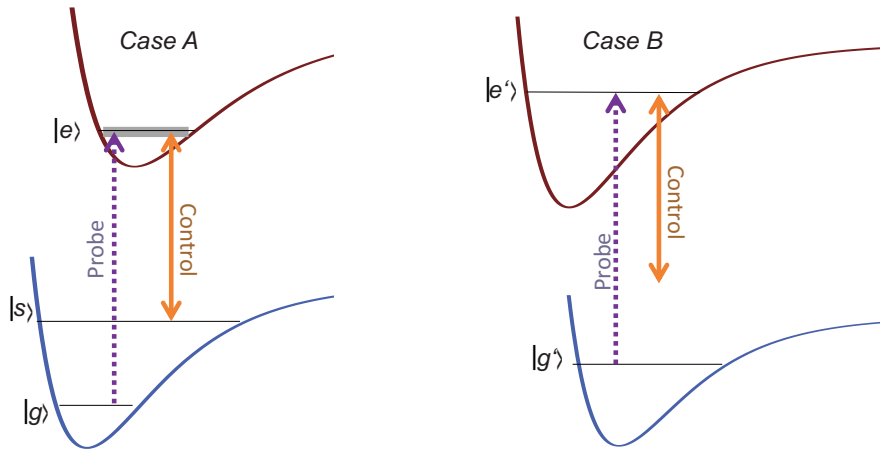}
\caption{(Color online) EIT assisted spectroscopic probing. A
strong control field eliminates one of the overlapping lines from
the absorption spectrum. This field is not in resonance with any
strong transition in the second molecule.} \label{figmodel}
\end{figure}



\vskip .6truein

\noindent
{\bf 2.2. Intermolecular EIT spectroscopy}

\vskip .2truein\noindent
As a first application we consider the use of EIT in resolving
spectral overlaps in the ``intermolecular'' case,
shown in Figure \ref{fig12}, in which one molecule
in a mixture overshadows the absorption of another.
Here EIT spectroscopy amounts to the application of a control
field which is in resonance with a transition belonging to just
one of the molecules/atoms in the mixture, thereby making that molecule
transparent at the probe's frequency,
bringing out the absorption lines of other molecules.
Such a configuration is
shown in Figure~\ref{figmodel}.

Denoting the absorption lines we wish to eliminate from the
spectrum as ``$A$ lines'', and the ones we want to preserve --
``$B$ lines'', the scheme will succeed if the control field
eliminating the $A$ lines will not modify the $B$ lines in any
substantial way. This condition is fulfilled, as discussed in
greater detail below, if the control field is far from resonance
for all the $B$ lines.

A simple illustration of the intermolecular scenario is given in Figure
(\ref{fig_one_eit}).
{Panel (a) shows two overlapping spectral lines of Lorentzian
profile, of the same height and same width $\G$, whose center-line
positions are separated by $\G/4$. This
small separation relative to the widths
does not allow resolving by ordinary means
one line in preference to the other. In Panel (b) we show
the result of a calculation performed according to
Eqs. (\ref{P_pr},\ref{dipole_AT}) with $\W_c=2.5\G,$
verifying that a
resonant control field effectively eliminates one of the $A$ lines, leaving,
due to its being off-resonance with the control laser field,
the $B$ line unaffected.
In order to estimate the required control intensity, we take a typical
value of $\G=0.01$ cm$^{-1}$. This translates into $\W_c=0.75$
GHz. For $\mu_{s,e}=0.1$ Debye,
the required control field intensity is merely $I_c=15$ kW/cm$^2$.

\vskip .2truein
\noindent
{\bf 2.3. Intramolecular EIT spectroscopy: Elimination of
many overlapping transitions in the same molecule}


\vskip .2truein\noindent In a similar fashion we can eliminate one
transition which overlaps with another within the same molecule.
This case may however be more complicated due to the fact that in
a thermal initial ensemble the spectrum may contain many
$|g\rangle$ to $|e\rangle$ transitions which strongly overlap.
\begin{figure}
\centering
\includegraphics[width=0.6\columnwidth]{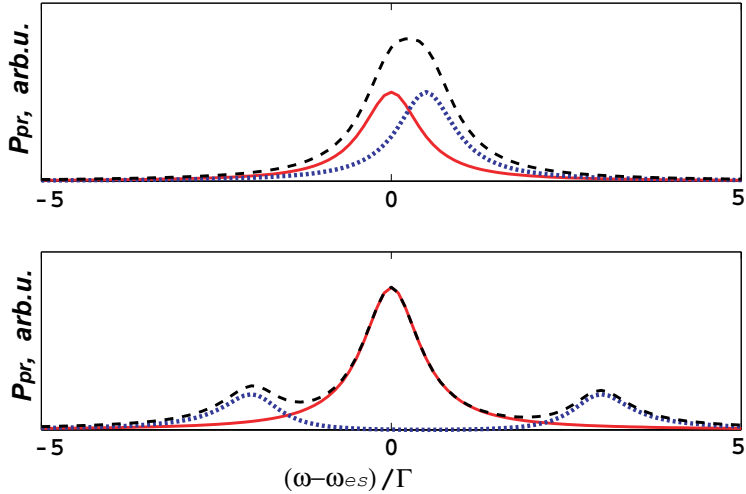}
\caption{(Color online) Use of EIT spectroscopy to resolve two
overlapping lines. The two overlapping absorption lines are shown
as dotted blue and solid red lines. Their joint envelope is shown
by the dashed black line. In the upper panel, the spectrum is
unperturbed. In the lower panel, one of the absorption lines has
been AT-split and removed from the spectrum.} \label{fig_one_eit}
\end{figure}
\begin{figure}
\centering
\includegraphics[width=0.8\columnwidth]{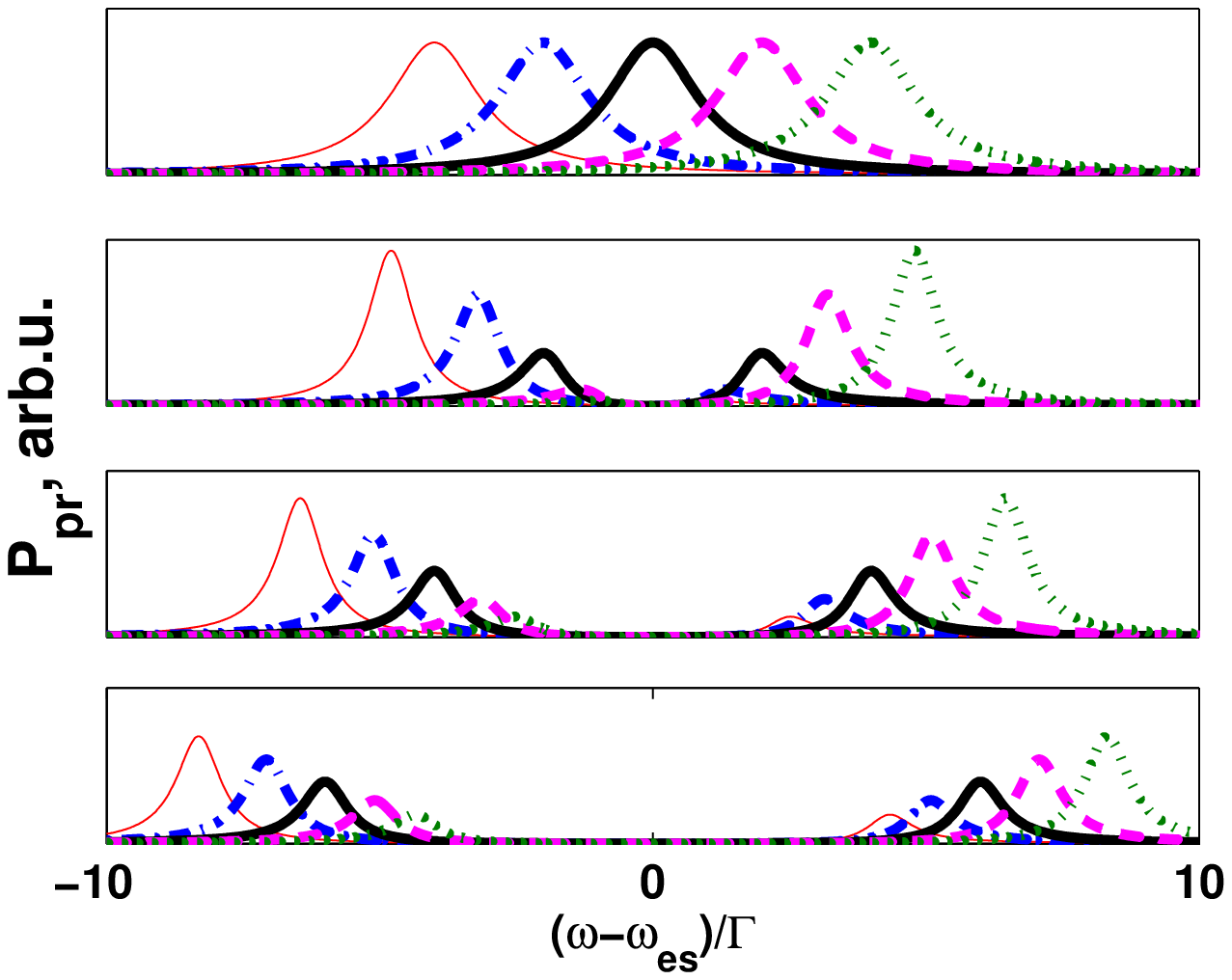}
\caption{(Color online). Few close lying transitions of the
control field which are close to the probe transition frequency
too. We demonstrate the behavior of the spectrum in several
control amplitudes. (a): $\Omega_{c}=0\G$; (b): $\Omega_{c}=2\G$;
(c): $\Omega_{c}=4\G$; (d): $\Omega_{c}=6\G$. } \label{figclr}
\end{figure}
Moreover, in large molecules, even if thermally cooled,
the $B$ line can be embedded in a congested
spectrum of close-lying transitions.
For the method to succeed,
one needs to ensure that {\it all} the neighboring
$A$ lines are made transparent or are pushed away,
not just those which are in exact resonance with the control field,
leaving no trace in the spectral region of
interest. The method will fail if some of the neighboring $A$
lines are pushed into the region of interest rather
than away from it.

The case of many $A$ absorption lines is investigated in
Fig.\ref{figclr}. Panel (a) shows a spectrum comprised of five
lines of the same $\G$ widths, whose line positions are
separated by 2$\G$. The task is to eliminate all
the spectral features from the region surrounding $\w=\w_{res}$.
Panels (b)-(d) show the effect of a control field, in resonance
with the central line, on all the lines in the spectrum. The
control Rabi frequency is set as $\Omega=2\G, 4\G$, and $6\G$
respectively. The change in the spectrum due to the control field,
which is now off-resonant for most of the lines, is calculated for
each line according to the formulas (\ref{P_pr},\ref{mu_nonres}).

Figure \ref{figclr} clearly shows that the control pulse can indeed part the
congested $A$ spectrum, removing $A$ lines that are both
on- and off-resonance with the control field. The
pulse intensities remain essentially the same as in the case of just two
overlapping lines.

\section*{3. CRITERIA OF APPLICABILITY}

\vskip .2truein\noindent
The condition for the  proposed method to work well
is for the control field to nullify or push away the $A$ lines absorption
while leaving the $B$ lines absorption intact.
One can
determine whether the $B$ lines are going to be
distorted  by calculating the energies of the eigenstates
modified due to the interaction with the control field. The new
frequencies arising in  the $B$ spectrum due to AT splitting
can be found via an {\it approximate formula} (\ref{ATenergies})
\begin{equation}\label{atmeasure}
E_{AT}=E_{e'}+\left[ \frac{\delta'}{2}\pm
 \left(
 \frac{\delta'}{2}+\frac{2\left|\Omega'_{c}\right|^{2}}{\delta'}\right)
\right]. \label{AT}
\end{equation}
Here $E_{AT}$ are the (dressed) energies of the AT doublets, $E_{e'}$
is the unperturbed energy of the $B$ $|e'\rangle$ state, $\delta'$ and
$\W'_c$ are the detuning and Rabi frequency for the strongest
transition involving the
$|e'\rangle$ state. The approximation is valid if
$\delta'\gg\Omega'_{c}$.

The parameters in Eq. (\ref{AT}) must be chosen in such a way
that the $B$ line is not distorted. If
one can choose the control field such that the detunings of all
the relevant transitions are high, and the corresponding dipole matrix
elements are
small, then  the $B$ absorption lines will be observed at their
natural positions.
%
Making an order-of-magnitude estimate, we see that if $\delta$
and $\W_c$ of the
$A$ lines satisfy the $\delta<\sim\G$, and $\W_c\simeq\G$ conditions,
the method will work if we can move the
$A$ lines by about a line width. Assuming that the $B$ Rabi frequencies
are of the same
order: $\W'_c\simeq\G$ (which is rather pessimistic because
the control frequency is off-resonance with respect to the $B$ absorption),
the requirement that the B line position not be moved by much means, using
Eq.(\ref{AT}), that
\begin{equation}
{\delta'} \gg {\G}~.%
\label{applicability}\end{equation}
The method will not work if  the molecule has a
dense spectrum near {\it both} the $|e\rangle$ {and} the $|s\rangle$
states, and if many of the $e$-$s$ transitions are strong.
In this case, due to the many AT-split $e$ levels
the $A$
spectrum may not part away sufficiently from the spectral region of interest.

\noindent
\section*{4. APPLICATIONS TO REAL MOLECULES \label{section-demonstration}}

We now present three realistic cases where we successfully implement
EIT spectroscopy.

\vskip .2truein\noindent
{\bf 4.1.  Two isotopomers of Cl$_2$.}

\vskip .2truein\noindent
As $B$ lines we choose transitions belonging
to the $^{35}{\rm Cl}^{37}{\rm Cl}$
minority isotopomer. As $A$ lines we choose transitions belonging to
the $^{35}{\rm Cl}_{2}$ majority isotopomer.
Table 1, obtained from Ref.\cite{Coxon1978}, shows the frequencies of
several $X^{1}\Sigma_{g}^{+} - B ^{3}\Pi_{0}^{+}$ transitions
of $^{35}{\rm Cl}_{2}$ and $^{35}{\rm Cl}^{37}{\rm Cl}.$
The table shows several pairs of ``intramolecular'' and ``intermolecular''
overlapping lines whose line-widths are typical of gas-phase conditions.
For example, the $^{35}{\rm Cl}_{2}$
$R_{v_{2-9}}^{J=59}$ transition and the $^{35}{\rm Cl}^{37}{\rm Cl}$,
$P_{v_{1-9}}^{J=28}$ transition
strongly overlap. (Here $P$ stands for a $\Delta J=-1$ transition,
$R$ - for a $\Delta J=1$ transition, with the subscript denoting the
vibrational quantum numbers involved in the transition.
$J$ designates the angular momentum of the lower state).

Since diatomic rotational energies can be approximated as
\begin{equation}
E_{rot} = B_vJ (J+1) + {\rm const},
\end{equation}
the separation between neighboring rotational
levels increases linearly with $J$. It therefore makes sense
to consider the case of high $J$ $A$ lines
because such lines are separated from
other rotational transitions. We have thus
chosen the $^{35}{\rm Cl}_{2}$ $R_{v_{2-9}}^{J=59}$
transition as our $A$ line, with
$|g\rangle$ being the $^{35}{\rm Cl}_{2}$
$X ^{1}\Sigma_{g}^{+}(J=59,v={2})$
state
and $|e\rangle$ being the $B ^{3}\Pi_{0}^{+}(J=60,v=9)$ state.
Given this $A$ line, the auxiliary $|s\rangle$ state was chosen to be
the
$^{35}{\rm Cl}_{2}$ $X ^{1}\Sigma_{g}^{+}(J=59,v={4})$ state,
giving rise to transition frequency
of $\w_{e,s} = 1.3286\cdot10^{4}$ cm$^{-1}$.

The control field was chosen to be detuned from the $B$ lines
of the $^{35}{\rm Cl}^{37}{\rm Cl}$ molecule by $\delta'=9$ cm$^{-1}$, with
the control Rabi frequency being $\W'_{c}=10^{-2}{\rm cm}^{-1}.$
The $B$ line shift, calculated according to Eq. (\ref{atmeasure}), of
$10^{-4}{\rm cm}^{-1},$
is negligible compared to the line width. Thus, the
control field leaves the
position and shape of the
$^{35}{\rm Cl}^{37}{\rm Cl}$, $P_{v_{1-9}}^{J=28}$
line essentially untouched, as required.
We conclude that EIT spectroscopy allows us to effectively eliminate
transitions of the majority $^{35}$Cl$_2$ isotopomer which overlap with
transitions associated with the $^{35}{\rm Cl}^{37}{\rm Cl}$ minority
isotopomer.

\begin{table}
\begin{center}
    \begin{tabular}{ | c | c | c | c |}
    \hline
    Isotopomer &
    $ ^{35,35}{\rm Cl}_{2}$ $v_{2-9}$ band    & $^{35,37}{\rm Cl}_{2}$ $v_{1-9}$ band      &$ ^{35,35}{\rm Cl}_{2}$ $v_{1-9}$ %
    band
    \\ \hline
     Rotational Branch,    & $R_{J=59}$,         & $P_{J=28}$,         & $P_{J=5}$,          %
     \\
     $\w_{ge}$ (cm$^{-1}$) & $1.312\cdot 10^4$   & $1.312\cdot  10^4$  & $1.312\cdot 10^4$   %
     \\ \hline
      Rotational Branch    & $-$                 & $R_{J=32}$,         &$R_{J=7}$,           %
      \\
      $\w_{ge}$ (cm$^{-1}$)&                     & $1.3122\cdot 10^4$  &  $1.322\cdot 10^4$  %
      \\ \hline
    Rotational Branch,     & $P_{J=55}$,         & $R_{J=33}$,         & $-$                 %
    \\
    $\w_{ge}$ (cm$^{-1}$)  & $ 1.3119\cdot 10^4$ &$ 1.3119\cdot 10^4$  &                     %
    \\
    \hline
    \end{tabular}
\end{center}
\caption{Overlapping lines in $^{35,35}{\rm Cl}_{2}$ and
$^{35,37}{\rm Cl}_{2}$. The letters $P$ and $R$ refer to the rotational
branches $\Delta J=\pm 1$. }
\end{table}


\vskip .2truein\noindent
{\bf 4.2. Eliminating two ``intramolecular'' overlaps in the
Chlorine molecule.}

\vskip .2truein\noindent
As surmised from Table 1, there is a strong overlap between the
$P_{v_{1-9}}^{J=5}$ and $R_{v_{2-9}}^{J=59}$ lines in
$^{35}{\rm Cl}_{2}$. Again we choose
to eliminate (an $A$ line) originating at a rather high $J$ quantum number.
A control field with the frequency $\w_c=1.3286\cdot10^{4}$ cm$^{-1}$
is used to couple the
$|e\rangle=B ^{3}\Pi_{0}^{+}(v=9,J=59)$ state to
the $|s\rangle=X ^{1}\Sigma_{g}^{+}(v=5,J=59)$ state.
As in the above, we first verify that the
parameters we choose for the control field are such as to not affect the
target ($B$-line) $=P_{v_{1-9}}^{J=5}$ transition.
In order to guarantee this, due to vibrational anharmonicity,
the detuning $\delta'$ in this case must be $\sim$20 cm$^{-1}$.
With this choice of parameters we can successfully ``push'' away the $A$ line
to reveal the (unaltered) $B$ line.

\begin{table} 
\begin{center}
    \begin{tabular}{ | l | l | l | p{5cm} |}
    \hline
      line & transition &  frequency, [cm$^{-1}$]   \\ \hline
     (a) & $P(1,10;22)^0 \leftarrow (0,9;23)^0 A$&  $231.14691 $ \\ \hline
     (b) & $Q(1,5;18)^0  \leftarrow (0,4;18)^0 E$ & $231.14924$  \\ \hline
     (c) & $Q(1,5;19)^0  \leftarrow (0,4;19)^0 E$ & $231.15178$  \\  \hline
     (d) & $Q(1,5;17)^0  \leftarrow (0,4;17)^0 E$ & $231.15353$  \\
    \hline
    \end{tabular}

\end{center}
\caption{Overlapping lines in ${\rm CH}_{3}{\rm OH}$\cite{G.1994}. }
\end{table}

\vskip .2truein\noindent
{\bf 4.3. Eliminating ``intramolecular'' overlaps in Methanol.}

\vskip .2truein\noindent
Compared to Chlorine, the spectrum of Methanol, ${\rm CH}_{3}{\rm OH}$,
is replete with intramolecular overlaps.
Table 2 lists several nearly overlapping frequencies of
rotational transitions \cite{G.1994,Methanol2}, where the line
notation is: the first entry refers to $P$, $Q$ or $R$
rotational branch; the next entry (the $(n,K,J)$ triad) is composed of,
$n,$ the torsional quantum number; $J$
the total angular momentum quantum number;
and $K,$ the projection of $J$ on the
molecular-axis quantum number. The
superscript $^0$ implies that there are no vibrational excitations
in any of the states, the transitions being purely torsional and rotational.
The last entry pertains to the ($A$ or $E$) point group irreducible
representation of both the $|e\rangle$ and
$|g\rangle$ states\cite{Methanol2}.

When substantial overlaps exist, it is not an easy matter
to derive numerically the individual transition line shapes from their joint
envelope\cite{G.1994,Methanol2}.
In contrast our EIT spectroscopy can solve this problem
directly: Using EIT to eliminate one of the lines from the
spectrum, and comparing the resulting spectral envelopes, one can
deduce the correspondent line shape.

The rotational transitions of Methanol are found in \cite{Methanol2}.
From the list, we see that a
$\w_c=249.291~{\rm cm}^{-1}$ control field can couple the $(1,3,19)$
$|e\rangle$ state to a $(0,4; 18)$ $|s\rangle$ state, and thus
eliminate the second transition in Table 2 from the spectrum.
At the same time, the control field is off resonance for
transitions form the other three excited states shown in Table 2.
The detuning for the $(0,4;19)$ and $(0,4;17)$ transitions is equal to 2
cm$^{-1}$. Repeating the arguments given above, we see
that the three remaining transitions are not influenced by a
control field of moderate intensity.

%

\section*{5. SUMMARY AND DISCUSSION}

We have introduced EIT spectroscopy and have shown that through it it is
possible to resolve overlapping spectral lines. The technique
uses the EIT effect to remove
unwanted transitions that mask some underlying transitions of interest.
We have discussed the range of applicability of the technique for both
sparse and congested spectra, and have demonstrated its viability
for Chlorine isotopomers and Methanol
molecules.

It is interesting to relate this technique to the use of Raman spectroscopy to
distinguishing between overlapping lines,
as shown in Fig.\ref{fig_cars}(a). We first
note that the AT effect can lead to saturation of the
Electronic Resonance Enhanced Coherent Anti-Stokes Raman
Scattering (ERE-CARS) spectroscopy. In ERE-CARS, the pump and
Stokes laser fields create a Raman coherence, driving a Raman
transition from the ground state $|g\rangle$ to the excited
vibrational state $\left|e\right\rangle$ of a molecule. The probe
field drives a resonant transition into an excited electronic
state $|s\rangle$, leading to emission of an anti-Stokes transition at
frequency $\w_{as}=E_s-E_g$. Early findings of Ref.
\cite{Chai2010} indicate that if the probe field is too strong, it
may cause Autler-Townes splitting of the $\left|e\right\rangle$ state, in the
way described by the equation (\ref{dipole_AT}), leading
to saturation of the ERE-CARS signal.

We note that the same effect can enable distinguishing overlapping
lines of different molecules. The implied scenario is similar to
those discussed Section 3. The scheme is shown in
Fig.\ref{fig_cars}(b) for the $A$ lines. The Pump and
Stokes fields create Raman coherences in the molecule. The CW
Probe, which may or may not be electronically resonant with an
electronic transition, measures the resulting emission at
anti-Stokes frequency of $\w_{as}=E_{e}+\w_{Probe}$.
Simultaneously with the Probe, one applies a strong Control field,
resonant with some electronic $A$ transitions and non-resonant
with the $B$ transitions. Thus for the $A$ transitions the $|e\rangle$
state is not populated, and the spectroscopic signal is
suppressed. The analytical description in this case is similar to
that given in Sections 2 and 3 of this paper.

\begin{figure}
\centering
\includegraphics[width=0.7\columnwidth]{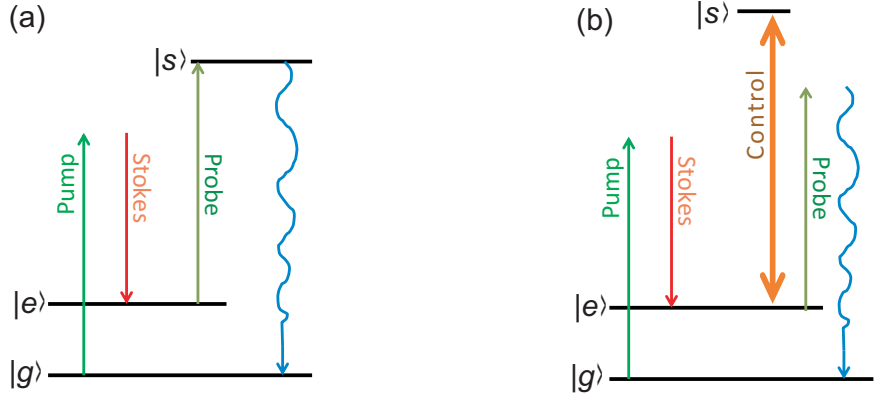}
\caption{(Color online). (a) ERE-CARS. The Pump and Stokes laser
pulses create a Raman coherence in a molecule. The Probe drives a
resonant transition into an excited electronic state $|s\rangle$,
leading to emission at the frequency $\w_{as}=E_s-E_g$. (b)
EIT-assisted CARS scheme. The Pump and Stokes pulses create a
Raman coherence, and the CW Probe drives emission at the
anti-Stokes frequency,  $\w_{as}=\w_e+\w_{Probe}$. The Control
field, applied simultaneously with the Probe, suppresses emission
at the frequency $\w_{as}$.
 }
\label{fig_cars}
\end{figure}

Our method can also be combined with a technique proposed in
Ref.\cite{Izmailov2008}, where it is pointed out that the
population in the transition eliminated via EIT ("case $A$") is
trapped in the dark state $|D\rangle\simeq |g\rangle$, while the
ground state $|g'\rangle$ of the $B$ transition is depleted.
A second probing, coming after the $B$ line is measured, will
find predominantly the $A$, rather than $B$, transition.
Thus within a single EIT arrangement one may be able to measure,
separately, each of the overlapping lines.

Further, EIT spectroscopy can be applied to measure a weak
spectral line superimposed on a strong and uncontrollable
background. In this case, one can make two measurements. In the
first one, the line is measured on top of the background. In the
second, a control field is applied to eliminate this line, and the
background alone is measured. Comparison of the two spectra will
yield the desired line shape, provided that the background is not
strongly influenced by the control field.


\centerline{{\bf Acknowledgments.}}

This work was supported by DTRA, NSERC, and by a Major Thematic Grant of UBC's
Peter Wall Institute for Advanced Studies.



\begin{thebibliography}{99}
\centerline{{\large\bf References}}

\vskip .2truein

\bibitem{Harris1997}
S. E. Harris,
\newblock Phys. Today, {\bf 50}, 36 (1997).

\bibitem{Boller1991}
K. -J. Boller, A. Imamoglu, and S. E. Harris,
\newblock Phys. Rev. Lett., {\bf 66}, 2593 (1991).

\bibitem{Field1991}
J. Field, K. Hahn, and S. E. Harris,
\newblock Phys. Rev. Lett., {\bf 67}, 3062 (1991).

\bibitem{fleishhauer05} M. Fleischhauer, A. Imamoglu, and J. P. Marangos,
                        Rev. Mod. Phys., {\bf 77}, 633 (2005).


\bibitem{Autler1955}
S. Autler, and C. Townes,
\newblock  Phys. Rev., {\bf 100}, 703 (1955).

\bibitem{Shapiro1972}
M. Shapiro,
\newblock J. Chem. Phys., {\bf 56}, 2582 (1972).

\bibitem{Fano1961}
U. Fano,
\newblock Phys. Rev., {\bf 124}, 1866 (1961).

\bibitem{G.Kurizki1989}
G. Kurizki, M. Shapiro, and P. Brumer
\newblock  Phys. Rev. B, {\bf 39}, 3435 (1989).


\bibitem{Shapiro2007}
M. Shapiro,
\newblock Phys. Rev. A., {\bf 75}, 013424 (2007).

\bibitem{Arimondo1976}
E. Arimondo and G. Orriols,
\newblock Lettere Al Nuovo Cimento, Series 2, {\bf 17}, 333 (1976).

\bibitem{Arimondo1996}
E. Arimondo, Prog. in Optics, {\bf 35}, 257 (1996).

\bibitem{N.Bloembergen1987}
N. Bloembergen
\newblock Pure and Applied Chemistry, {\bf 55}, 1229 (1987).

\bibitem{Clark1988} R. J. H. Clark, {\it Advances in Non-Linear Spectroscopy (Advances in Spectroscopy}
1th edition, (John Wiley \& Sons, New York, 1988).

\bibitem{Brumer86} P. Brumer and M. Shapiro, Chem. Phys. Lett.,
                   {\bf 126}, 541 (1986).

\bibitem{Lukin1997}
M. D. Lukin, M. Fleischhauer, A. Zibrov, H. Robinson, V. Velichansky, L. Hollberg, and M. Scully,
\newblock  Phys. Rev. Lett., {\bf 79}, 2959 (1997).

\bibitem{Hau1999}
L. V. Hau, S. E. Harris, Z. Dutton, and C. H. Behroozi,
Nature, {\bf 397}, 594 (1999).

\bibitem{Hau2001}
Ch. Liu, Z. Dutton, C. H. Behroozi and L. V. Hau
Nature, {\bf 409}, 490 (2001)


\bibitem{Walsworth2001}
D. F. Phillips, A. Fleischhauer, A. Mair,
R. L. Walsworth, and M. D. Lukin. Phys. Rev. Lett., {\bf 86}, 783 (2001).

\bibitem{Matsko2001}
A.~Matsko, Y.~Rostovtsev, O.~Kocharovskaya, A.~Zibrov, and M.~Scully.
\newblock {Phys. Rev. A}, {\bf 64}, 043809 (2001).

\bibitem{Dey2003}
T. Dey and G.~Agarwal.
\newblock {Phys. Rev. A}, {\bf 67}, 033813 (2003).

\bibitem{Fleischhauer2002}
M.~Fleischhauer and M.~D. Lukin.
\newblock {Phys. Rev. A}, {\bf 65}, 022314 (2002).

\bibitem{Duan2001}
L. M. Duan, M. D. Lukin, J. I. Cirac, and P. Zoller,
\newblock Nature, {\bf 414}, 413 (2001).

\bibitem{Schwindt2004}
P. D. D. Schwindt, S. Knappe, V. Shah, L. Hollberg,
J. Kitching, L.-A. Liew, and J. Moreland,
App. Phys. Lett., {\bf 85}, 6409 (2004).

\bibitem{Affolderbach2005}
C. Affolderbach, C. Andreeva, S. Cartaleva, T. Karaulanov, G. Mileti, and D. Slavov,
App. Phys. B. {\bf 80}, 841 (2005).


\bibitem{Taichenachev1999}
A. Taichenachev, A. Tumaikin, and V. Yudin,
\newblock Phys. Rev. A., {\bf 61}, 011802 (1999).

\bibitem{Lezama1999}
A. Lezama, S. Barreiro, and A. Akulshin,
\newblock Phys. Rev. A., {\bf 59}, 4732 (1999).

\bibitem{lics} Y. I. Heller and A. K. Popov Opt. Commun., {\bf 18}, 449 (1976).
\bibitem{knight} P. E. Coleman, P. L. Knight, and K. Burnett, Opt. Commun.,
                   {\bf 42}, 171 (1982);
P. L. Knight, M. A. Lauder, and B. J. Dalton, Phys. Rep.,
                   {\bf 190}, 1 (1990).
\bibitem{charalambidis} S. Cavalieri, R. Eramo, L. Fini, M. Materazzi,
O. Faucher, and D. Charalambidis, Phys. Rev. A, {\bf 57}, 2915,  (1998).
\bibitem{halfmann} T. Halfmann, L. P. Yatsenko, M. Shapiro, B. W. Shore and
K. Bergmann, Phys. Rev. A, {\bf 58}, R46 (1998).

\bibitem{Goren2004}
C. Goren, A. Wilson-Gordon, M. Rosenbluh, and H. Friedmann,
\newblock Phys. Rev. A., {\bf 69}, 063802 (2004).

\bibitem{Hessel1974}
E. W. Smith, M. M. Hessel, and R. E. Drullinger
\newblock  Phys. Rev. Lett., {\bf 33}, 1251 (1974).

\bibitem{Caldwell1980}
C. D. Caldwell, F. Engelke, and H. Hage,
\newblock Chem. Phys., {\bf 54}, 21 (1980).

\bibitem{Coxon1978}
J. A. Coxon and R. Shanker,
\newblock J. Mol. Spectry, {\bf 69}, 109 (1978).

\bibitem{G.1994}
G. Moruzzi and H. Xu,
\newblock J. Mol. Spectry, {\bf 165}, 233 (1994).


\bibitem{Shapiro2002}
M. Shapiro and P. Brumer,
\newblock {\it Principles of the quantum control of molecular processes}
(John Wiley \& Sons, New York, 2003)

\bibitem{Izmailov2008}
A. Ch. Izmailov,
\newblock Laser Physics, {\bf 18}, 855 (2008).



\bibitem{Chai2010}
N. Chai, R. P. Lucht, W. D. Kulatilaka, S. Roy, and J. R. Gord,
\newblock J. Chem. Phys., {\bf 133}, 084310 (2010).



%
%
\bibitem{Methanol2}
G. Moruzzi, F. Struma, J. C. S. Moraes {\it et al.,~}%
J. Mol. Spectry, {\bf 153}, 511 (1992).

\end{thebibliography}

\end{document}